# About using of a compact supermirror transmission polarizer in the neutron research facilities of the PIK reactor

V.G. Syromyatnikov

1 - *Petersburg Nuclear Physics Institute (PNPI), National Research Center "Kurchatov Institute", Gatchina, Russia*
2 - *Department of Physics, Saint Petersburg State University, St. Petersburg, Russia*

E-mail address: syromyatnikov_vg@pnpi.nrcki.ru

## Abstract

The prospects of using last version of a compact supermirror transmission polarizer *TRUNPOSS* on silicon in modern neutron research facilities of the instrumental base being created for the PIK reactor (PNPI, Gatchina, Russia) will discuss in the paper. The results of calculations of the parameters and main characteristics of this polarizer for using it in *IN2, IN3, SEM, DEDM, TENZOR* facilities are discussed in detail.

## 1. Introduction.

Recently, transmission neutron supermirror polarizers have become widely used in the neutron experiment using full neutron polarization analysis. One of the main advantages of such polarizers is that the beam coming out of the polarizer has the same direction as the beam at the entrance. This makes it possible to quickly and simply move from measurements with polarized neutrons to measurements with non-polarized ones, practically without rebuilding the facility. Well known neutron transmission supermirror polarizers: the most widely used a V-cavity [1-3] and solid-state devices: a transmission bender with a collimator [4], an S-shaped bender [5] and a kink polarizer [6, 7].

The main parameters of these polarizers are briefly described in [8].

Recently, in [9] and [10], transmission solid-state polarizing bender was considered for three-axis spectrometer and for analyzer device for MIEZE, respectively.

The kink polarizer has a significantly shorter length and a cheaper supermirror polarizing coating with a lower *m* parameter compared to V-cavity. The disadvantage of the kink polarizer is the insufficiently wide angular range of the beam, in which the beam coming out of the polarizer is highly polarized and the need to use a Soller collimator at the exit of the polarizer.

In [8], a kink polarizer is considered, in which the angular range of the output beam with high polarization is significantly increased due to the use of an additional element – a straight polarizing supermirror neutron guide installed at the exit of the kink polarizer. This compact polarizer operates in small magnetic fields. In this polarizer, the remanent properties of polarizing supermirrors were used. In this case, it is required to magnetize the parts of the polarizer in mutually opposite directions, which presents certain difficulties and requires additional research.

A new version of the neutron multichannel transmission solid-state supermirror polarizer, in which both its parts: the kink polarizer and the straight polarizing neutron guide are located in saturating magnetic fields, is considered in this paper. This polarizer, as well as its variant with the kink and neutron guide remanent states, is designed to polarize large-area beams with a wide angular distribution and to analyze the polarization of such beams scattered on the sample. Examples of the use of this polarizer in neutron physics facilities of the new high-flux research reactor PIK (NRC KI-PNPI) are considered. The material of this paper was presented as oral contributions to Asia-Oceania Conference on Neutron Scattering (AOCNS-2023) (December 2–8, 2023, Dongguan, China) and to Sino-Russia meeting on frontiers of neutron scattering (SRNS-2024) (October 8-11, 2024, Ekaterinburg, Russia).

First, let's turn to the design and parameters of the kink polarizer.

## 2. Kink polarizer

A scheme of a neutron kink polarizer is shown in Fig. 1 [6, 7]. This polarizer is much more compact than the V-cavity!

The polarizer consists of $N$ broken channels [7] sandwiched between two holders. The number of broken channels and their width are determined by the required cross-section of the beam used in the facility. $N$ such channels are used, which are pressed against each other without air gaps. A divergent neutron beam enters the input of each channel.

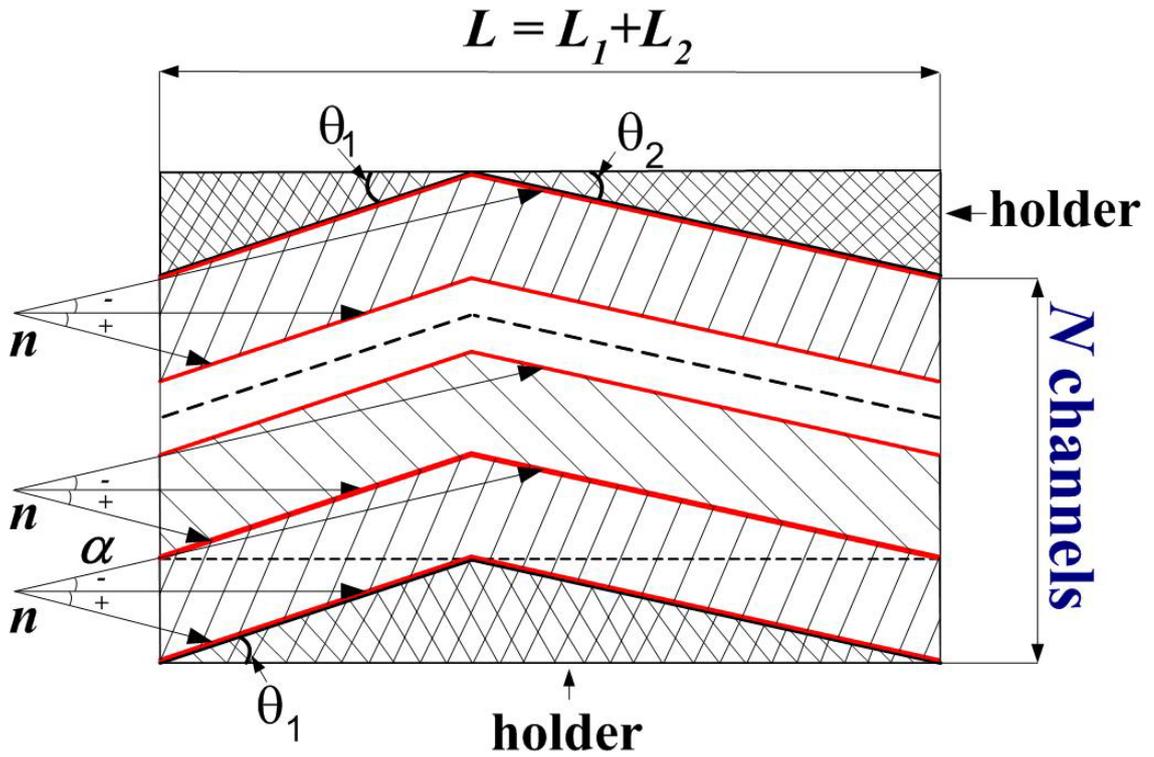

**Fig. 1.** A scheme of the neutron kink polarizer. General view from above.

The scheme of one channel is shown in Fig. 2. Channel parameters: angles $\theta_1$ and $\theta_2$ relative to the horizon line, lengths $L_1$ and $L_2$. The channel can be symmetrical. In this case, $\theta_1 = \theta_2$ and $L_1 = L_2$. The polarizing supermirror coating SM (highlighted in red) covers the polished surfaces of the channels with a width of $d$. M is a medium for neutron propagation. M is a material with a small neutron absorption cross-section (silicon, quartz, sapphire, etc.).

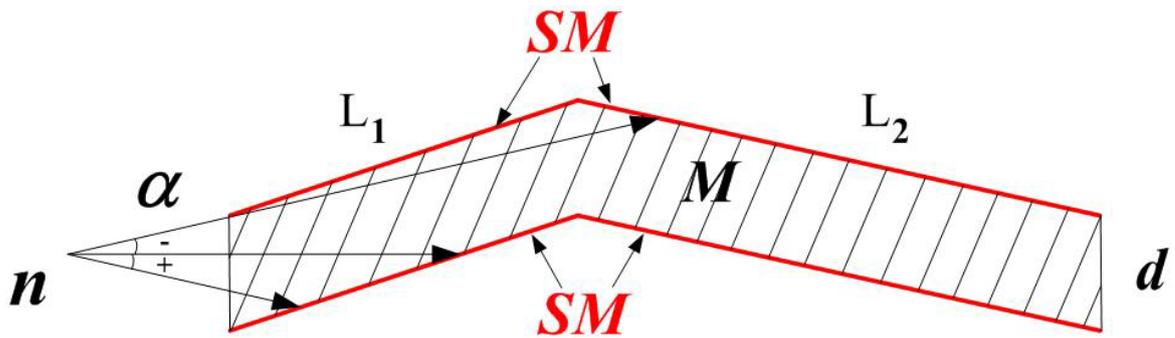

**Fig. 2.** A scheme of one channel of the neutron kink polarizer. General view from above.

For the (-) spin component of the beam, the neutron-optical potentials of the layers of the supermirror coating and the material of plate must be close to each other so that the critical angle is close to zero for the "material - supermirror" boundary. Therefore, the (-) spin component of the beam is practically not reflected from this boundary.

For the (+) spin component of the beam, the neutron-optical potentials of its layers must differ significantly from each other, so that the corresponding critical angle is large for the same boundary. In this case, the reflection from the "SM – M" boundary will be significant. Variants: SM - Fe/Si, CoFe/TiZr; M - silicon.

The polarizer plates are located in the field of the magnetic polarizer system. This field lies in the plane of the supermirror and is directed perpendicular to the plane of the drawing. The magnetic layers of the supermirrors are magnetized to saturation by the applied field.

The polarizing CoFe/TiZr ($m = 2$) supermirror [11] PNPI was used in the simulation of the polarizer.

Figure 3 shows the dependences of the reflectivity $R^+$ and $R^-$ from momentum transfer for both spin components of the beam in the saturating magnetic field for this supermirror. The curves for this supermirror were used in the calculations of the characteristics of the polarizer. It follows from the graphs that the spectral polarizing efficiency of the supermirror $P$ is very high and close to 1, since $P = \frac{R^+ - R^-}{R^+ + R^-}$, and $R^+ \gg R^-$! $R^+$ corresponds to the case when the neutron spins are parallel to the induction vector of the magnetic layers of the supermirror, and $R^-$ when these vectors are antiparallel.

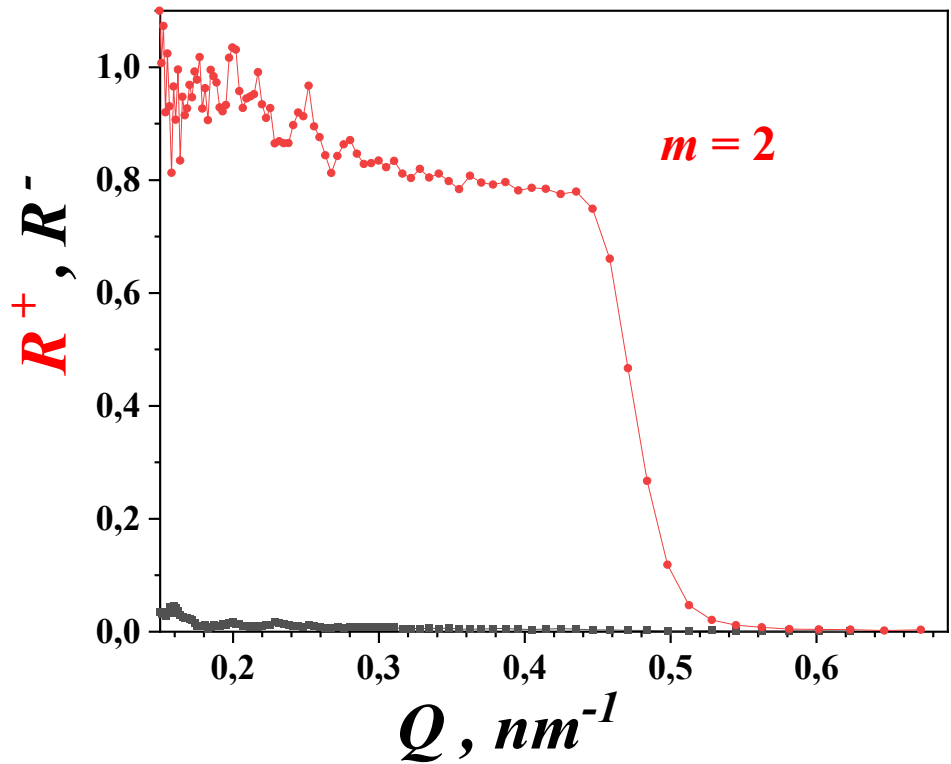

**Fig. 3.** The dependences of the reflectivity $R^+$ and $R^-$ from momentum transfer for both spin components of the beam in the saturating magnetic field for polarizing CoFe/TiZr ($m = 2$) supermirror PNPI.

The scheme of the kink consisting of two separate parts (shoulders) is shown in Fig. 4. This scheme is slightly modified compared to the scheme in Fig. 1. Each part is a stack of $N$ silicon wafers with lengths $L_1$ and $L_2$, respectively, pressed against each other without air gaps on both sides. Each stack is rotated relative to the horizon by an angle of $\theta_1$ and $\theta_2$, respectively. The thicknesses of the plates are the same and equal to $d$. A polarizing CoFe/TiZr supermirror coating with parameter $m = 2$ is evaporated to each side of the plate. The kink parameters are presented in Table 1. To calculate the transmission of a neutron flux through such a kink, a program prepared in the McStas environment [12] was used in accordance with the scheme in Fig. 4.

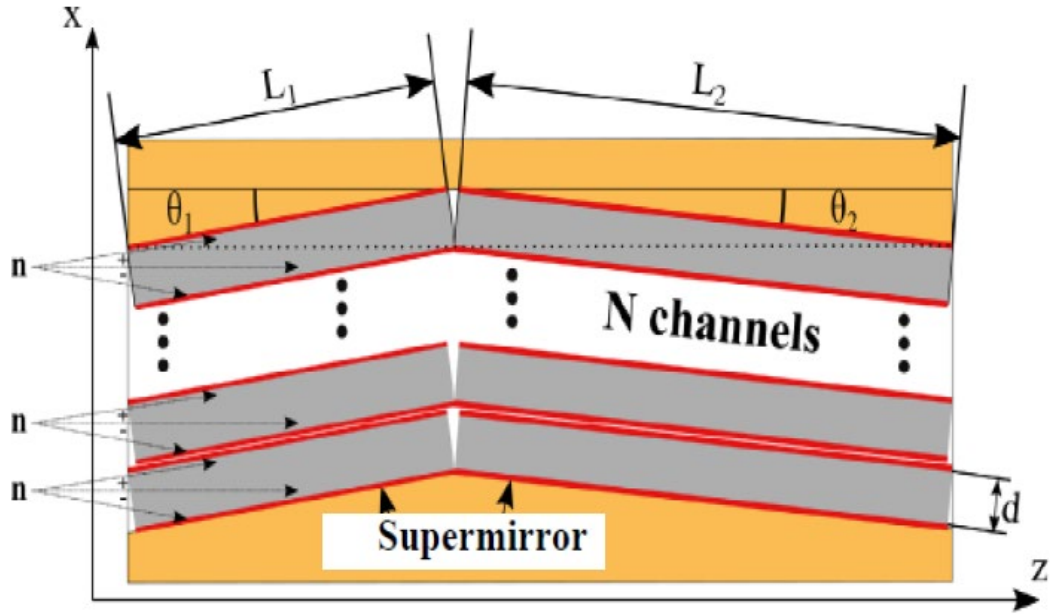

**Fig. 4.** The scheme of kink polarizer consisting of two separate parts (shoulders).

**Table 1.** Parameters of the neutron kink polarizer.

| $d$, mm | $L_1$, mm | $L_2$, mm | $\theta_1$, mrad | $\theta_2$, mrad | $m$ |
|---|---|---|---|---|---|
| 0.30 | 41 | 64 | 7.4 | 4.7 | 2 |

The results of calculations of transmission of a neutron beam with a wavelength of 4 Å through a kink polarizer depending on the angle for both spin components of the beam are shown in Fig. 5. The black line represents the angular distribution of the beam intensity at the entrance to the polarizer with a divergence of ± 0.45 degrees. The curve of the angular intensity distribution at the kink exit for the (+) spin component of the beam is shown in red. The angular intensity distribution of the (+) spin component of the beam at the kink exit contains a set of noticeable peaks due to the reflection of neutrons of this component from supermirror coatings. At the same time, there is a dip in intensity near the axis of the incident beam. The angular intensity distribution of the (-) spin component of the beam at the exit of the kink, as follows from the figure, coincides with the input angular distribution, since absorption in silicon was no taken into account here and neutron reflection of the (-) spin component of the beam from supermirror coatings was neglected.

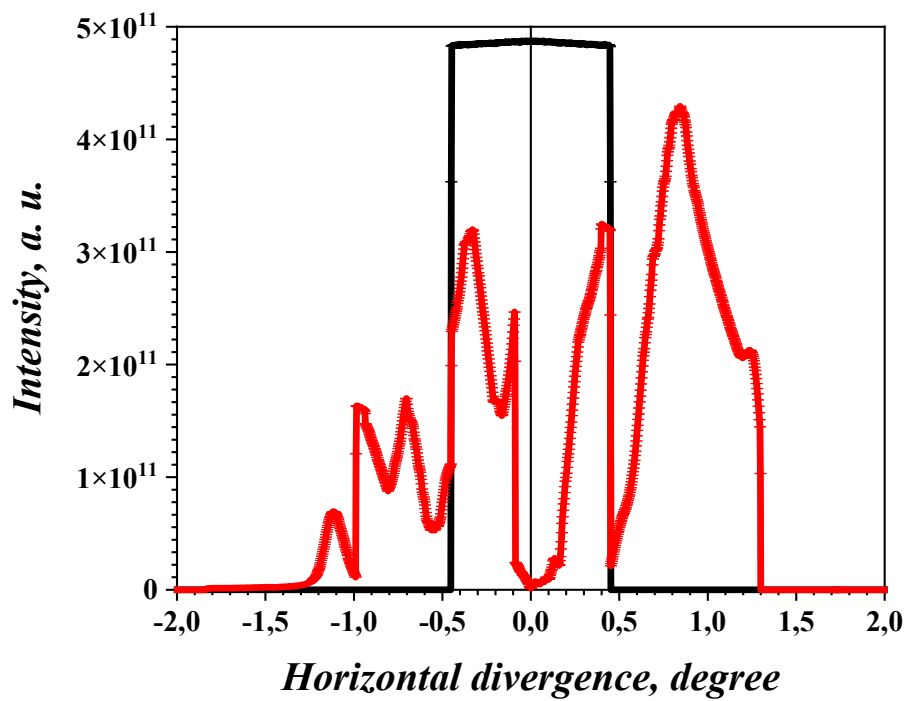

**Fig. 5.** The calculated transmission of the intensity of a neutron beam with a wavelength of 4 Å through a kink polarizer, depending on the angle for both spin components of the beam. Here, the intensity at the entrance of the polarizer and the (-) component of the beam at the exit of the polarizer are shown in black, the intensity of the (+) component of the beam at the exit of the polarizer is shown in red.

The same curves as in Fig. 5, but on a smaller scale, are shown in Fig. 6. All peaks of the (+) spin component of the beam must be removed from the exit beam. To do this, you will need to install a Soller collimator with "black" walls at the exit of the polarizer. This collimator must pass a beam with a divergence not exceeding ± 0.1 degrees in order to pass the neutrons of the (-) spin component of the beam and at the same time remove all peaks of the (+) spin component of the beam from the beam. Thus, at the exit from the kink, the beam will have a high level of negative polarization. Unfortunately, the angular width of this beam, as follows from Fig. 6, will be almost 5 times less than the beam width at the entrance, i.e. the beam transmission through the kink will not be high enough.

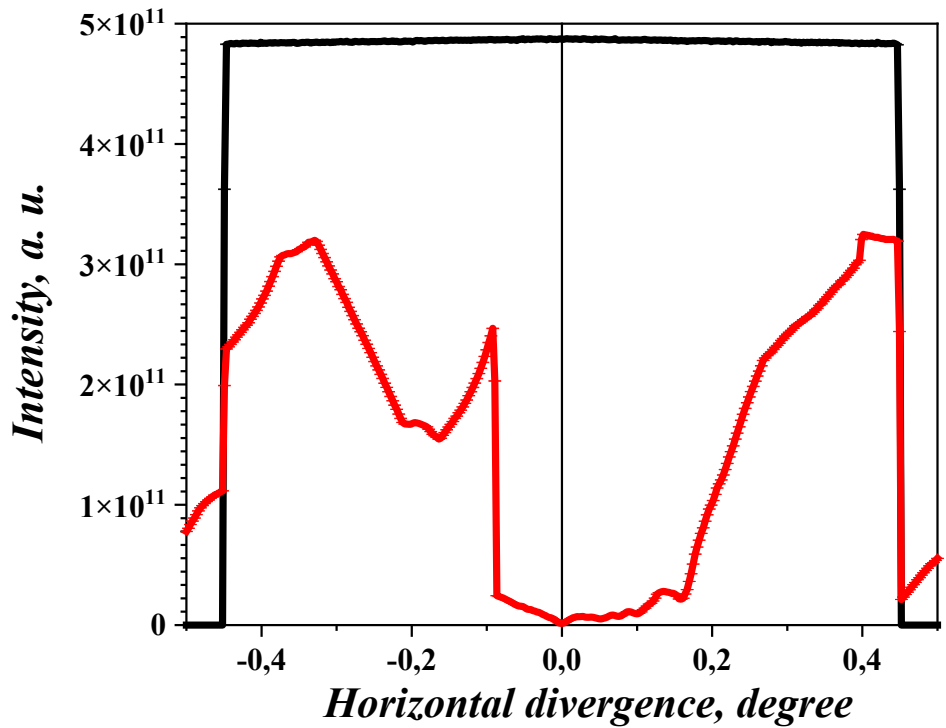

**Fig. 6.** The same curves as in Fig. 5, but in a smaller angular range.

### 3. A new version of the compact transmission polarizer with a kink part.

Another variant of the transmission polarizer, along with the considered multichannel transmission polarizer, which uses the remanent properties of a polarizing supermirror, is proposed in [8]. In this polarizer, both its parts (the kink or solid-state bender and the straight neutron guide) are in saturating magnetic fields, while between these elements there is a constantly switched-on spin flipper (or π-rotator) (Fig. 7). As a spin-flipper option, you can consider the Mezei spin-flipper in a constantly switched-on state, because this spin-flipper has compact dimensions. In both versions of the polarizer, the angular intensity distribution range of the output beam with high polarization is increased compared to the above-mentioned kink polarizer. The kink polarizer and the straight polarizing neutron guide have polarizing supermirror coatings evaporated on both sides of each silicon substrate in both versions of polarizer. At the same time, for neutron guide substrates, a neutron-absorbing sublayer (for example, TiZrGd) is additionally evaporated on top of the supermirror coating.

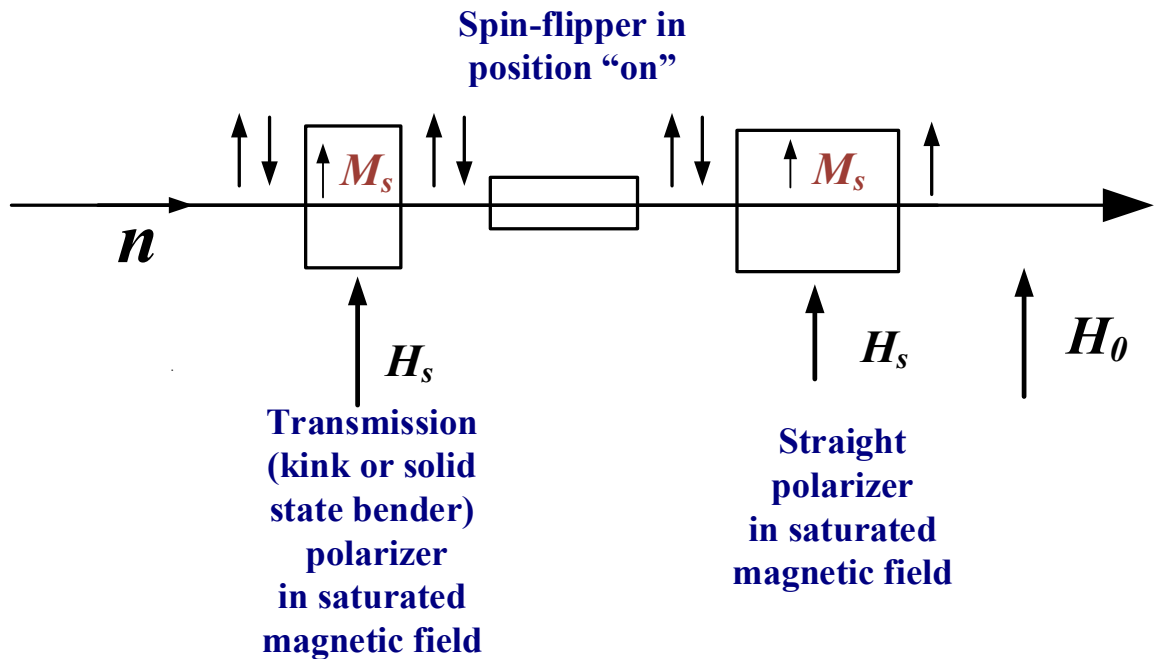

**Fig. 7.** A new version of the transmission neutron polarizer with a kink part [8].

Let's consider separately the transmission of neutrons of each spin component of the beam through the polarizer shown in Fig. 7, using Fig. 8.

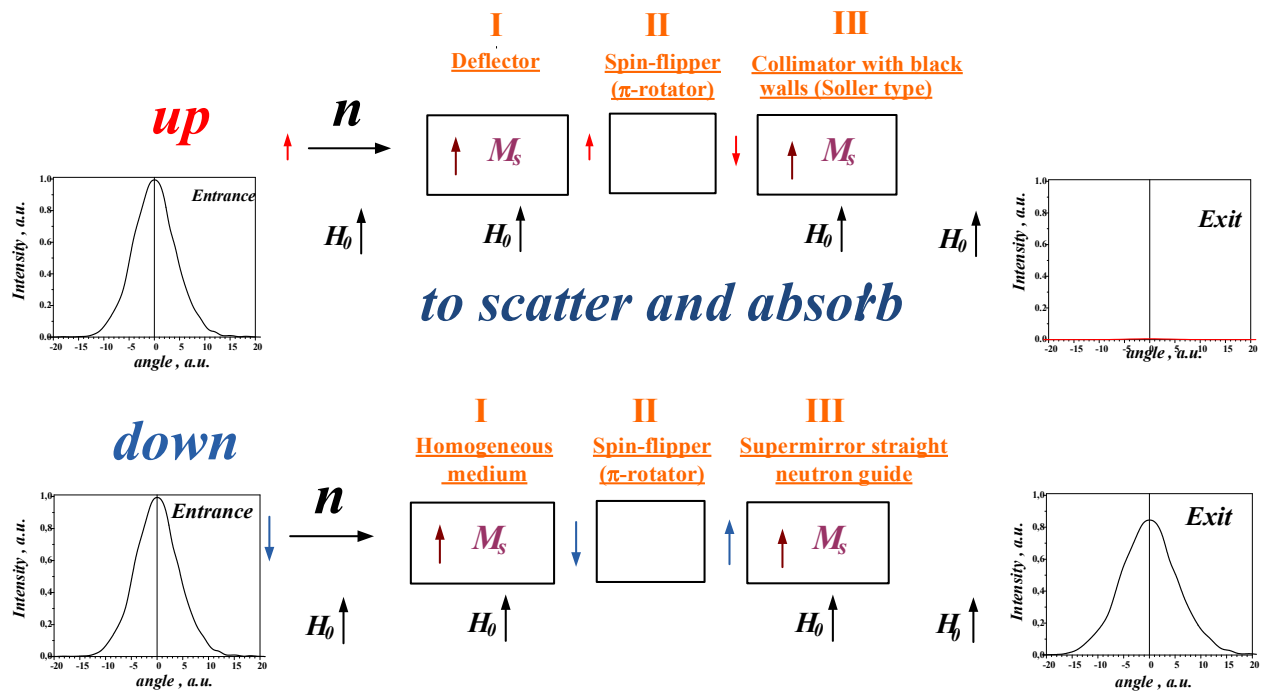

**Fig. 8.** The scheme of transmission of both spin components of the beam through a new transmission neutron polarizer with a kink part.

Let's start with the (+) spin component of the beam. The I-th part of the kink polarizer for the (+) spin component of the beam is a deflector, since it significantly deflects the neutrons of this component from their original trajectories. At the same time, the beam divergence increases significantly, and in the area of angles close to the beam axis it drops to almost zero. There is a spin flipper behind the kink (Part II). It is always in the switched-on state, i.e. it reverses the spins of neutrons released from the I-th part of the kink polarizer. When passing through a straight polarizing neutron guide (Part III), neutrons of the inverted (+) spin component of the beam are reflected from the walls with a very low reflection coefficient, since the orientations of neutron spins and magnetization vectors of the magnetic layers of supermirrors are antiparallel. In fact, in this case, part III operates like a standard Soller collimator. It would seem that neutrons should pass through it, because it has direct line-of-sight in the appropriate angular range. But there are simply no neutrons in this angle range, because they were scattered in the I-th part of the polarizer and then, practically without reflecting from the walls of the straight neutron guide, they are absorbed in its absorbing sublayer. As a result, there are practically no neutrons (+) spin component of the beam at the exit of the polarizer. Thus, for the (+) spin component of the beam, the polarizer performs two functions: <u>scatter and absorb</u>!

For neutrons of the (-) spin component of the beam, the situation is different. The Ist part of the polarizer (kink) for them is a homogeneous medium, which they pass without deviations from their original trajectories, only slightly reducing their intensity due to absorption in the material of plate. This is due to the fact that the potentials of the supermirror layers and the substrate material for the (-) spin component of the beam are close. After passing through the kink, the neutron spins are flipped by a spin flipper. Further, these neutrons pass through the straight neutron guide reflecting from its walls with a high reflection coefficient, because in this case, the neutron spins and the vector of magnetization of the magnetic layers of the supermirror are parallel to each other. As is known, the beam does not increase its divergence, when passing through a straight neutron guide. As a result, at the exit from the polarizer, the angular beam profile of this spin component repeats the beam profile at the entrance to the polarizer with a slight attenuation due to absorption in the substrate material. Thus, the beam passing through the polarizer will have a high level of negative polarization.

A scheme (top view) of the proposed model of a multichannel transmission polarizer consisting of three parts: a kink, a spin-flipper (or a $\pi$-rotator) and a straight polarizing neutron guide is shown in Fig. 9. The kink and the neutron guide are in saturating magnetic fields. The field for parts I and III is directed upwards perpendicular to the plane of the drawing, so that these parts of the polarizer have the same direction of magnetization. Each channel of the kink and neutron guide is a double-sided polished silicon wafer.

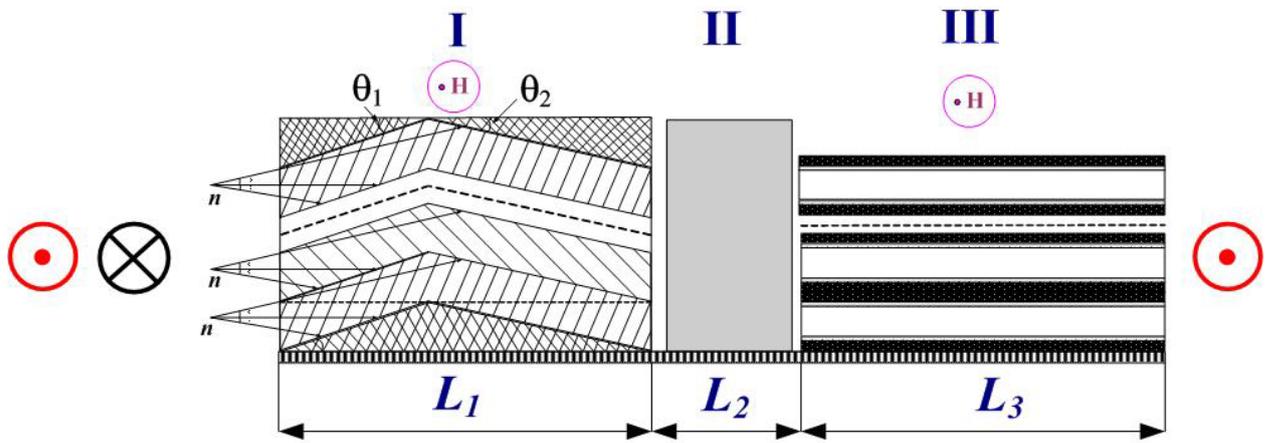

I – kink polarizer
II - spin-flipper (π-rotator)
III – straight polarized neutron guide

**Fig. 9.** The scheme (top view) of the proposed model of a multichannel transmission polarizer consisting of three parts: a kink, a spin flipper (or a π-rotator) and a straight polarizing neutron guide.

A polarizing supermirror CoFe/TiZr ($m = 2$) coating is evaporated on each side of the plate (Fig. 10). For the neutron guide channel, an additional absorbing TiZrGd coating is evaporated on top of this coating (Fig. 11).

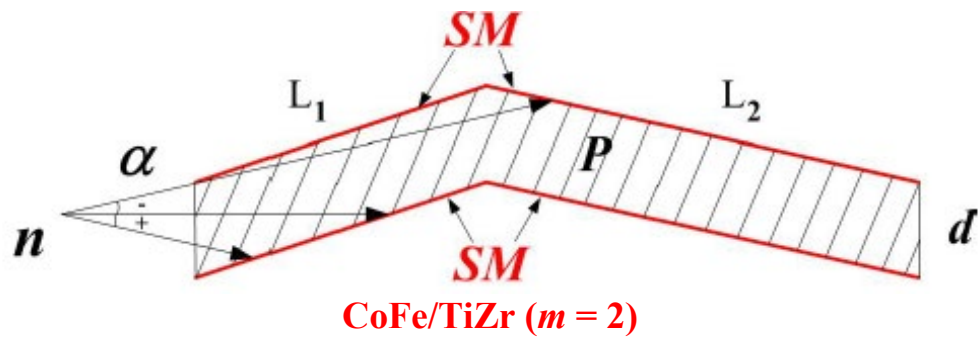

**CoFe/TiZr ($m = 2$)**

**Fig. 10.** The scheme of one kink channel for the proposed model of a transmission polarizer.

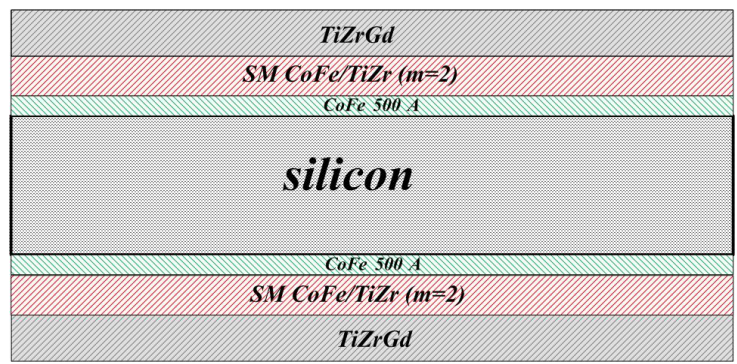

**Fig. 11.** The scheme of a single channel of a straight polarizing neutron guide for the proposed model of a transmission polarizer.

The proposed name for this polarizer is **TRUNPOSS** (**Tr**ansmission **Ru**ssian **N**eutron **Po**larized **S**upermirror **S**ystem). This name can also be applied to a polarizer that uses the remanent properties of polarizing supermirrors [8].

# 4. Proposals for the use of a new polarizer at the neutron physics facilities of the PIK reactor.

A proposal for the use of a new polarizer at few neutron physics facilities of the PIK reactor was considered.

The scheme of the *IN2* inelastic neutron scattering spectrometer in the polarization mode is shown in Fig. 12. The main parameters of the *IN2* beam are: the wavelength range is 2.4 – 6 Å, the beam cross section is significant - 30x116 $mm^2$, while the length of the polarizer should not exceed 300 mm. On *IN2*, it is planned to use this polarizer as a polarizer and analyzer (items 8 and 19, correspondingly).

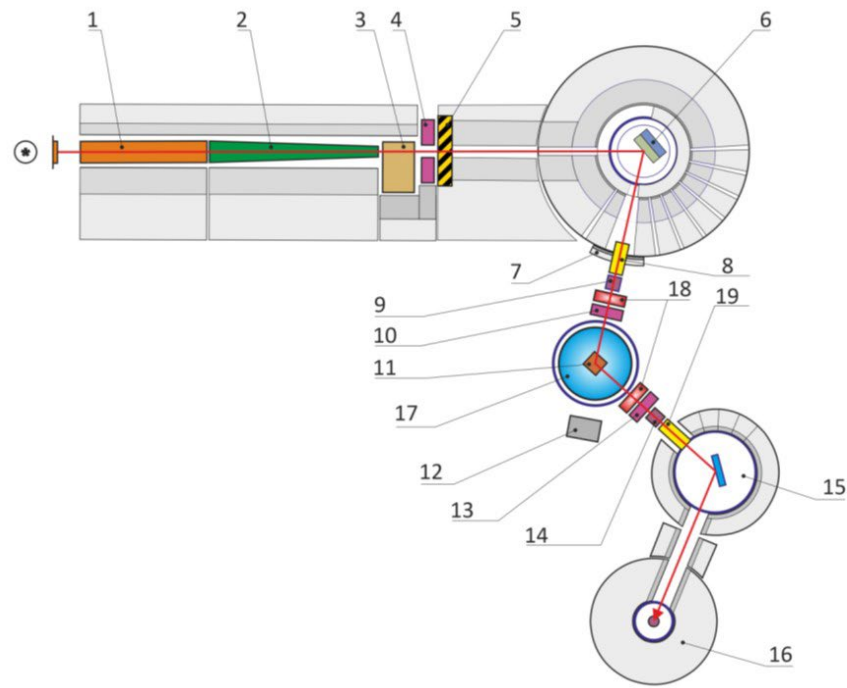

**Fig. 12.** The scheme of the *IN2* inelastic neutron scattering spectrometer in the polarization mode.

The scheme of a facility for measuring the electric dipole moment of a neutron using the crystal diffraction method (*DEDM*) is shown in Fig. 13. The main parameters of the *DEDM* beam are: wavelength 5 Å, beam cross section is significant - 100x100 $mm^2$. *DEDM* plans to use this polarizer to polarize the beam incident on the sample (item 7).

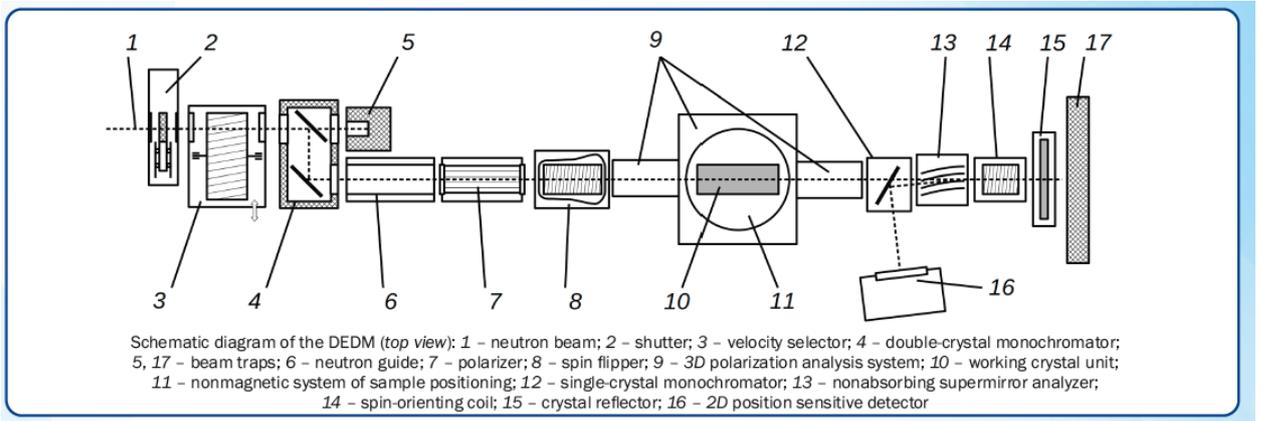

**Fig. 13.** The scheme of a facility for measuring the electric dipole moment of a neutron using the crystal diffraction method (*DEDM*).

The scheme of facility of the *SEM* - neutron spin-echo spectrometer in the configuration of intensity modulation with zero impact *MIEZE* (**M**odulation of **IntE**nsity with **Z**ero **E**ffort ) is shown in Fig. 14. The main parameters of the *SEM* beam are: wavelength range 2.4 - 12 Å, beam cross section - 30x30 $mm^2$. This polarizer is planned to use as an analyzer (A) at *SEM* facility.

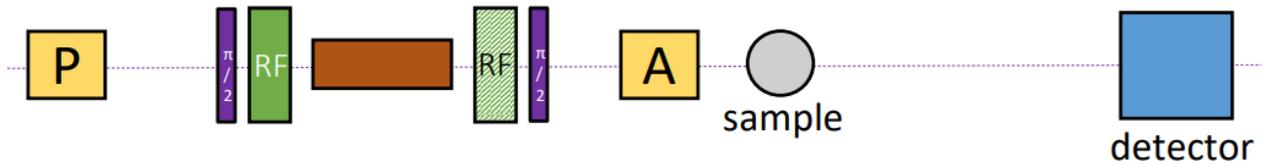

**Fig. 14.** The scheme of *SEM* facility - neutron spin-echo spectrometer in the configuration of intensity modulation with zero impact *MIEZE*.

The parameters of the polarizer (Fig. 9) for *IN2, DEDM, SEM and TENZOR* facilities are shown in Table 2.

**Table 2.** The parameter of polarizer (Fig. 9) for *IN2, DEDM, SEM* and *TENZOR* facilities.

| $d$, mm | $L_1$, mm | $L_2$, mm | $L_3$, mm | $\theta_1$, mrad | $\theta_2$, mrad | $m$ |
|---|---|---|---|---|---|---|
| 0.30 | 41+64 | 50 | 120 | 7.4 | 4.7 | 2 |
| 0.15 | 21+32 | 50 | 60 | 7.4 | 4.7 | 2 |

The calculated graphs of angular intensity distributions for the beam at the entrance (black curve) with a divergence of ± 0.45 degrees, for (+) spin component of the beam at the exit from the kink (red curve) and for (+) spin component of the beam at the exit from the straight polarizing neutron guide (blue curve) for a wavelength of 4 Å are shown in Fig. 15 for a polarizer with the parameters from Table 2.

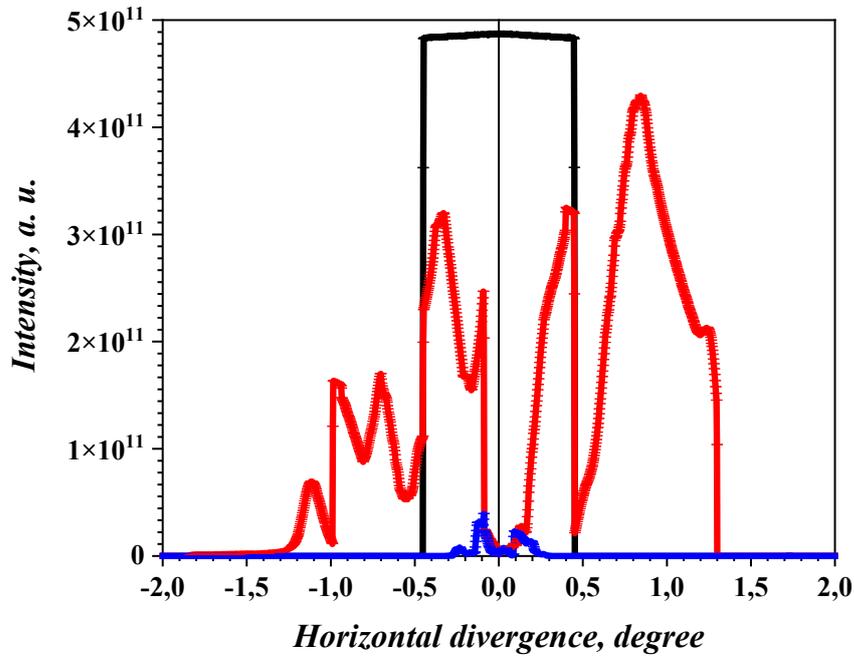

**Fig. 15.** The calculated graphs of angular intensity distributions for the beam at the entrance to the polarizer (black curve) with a divergence of ± 0.45 degrees, for (+) spin component of the beam at the exit from the kink (red curve) and for (+) spin component of the beam at the exit from the straight polarizing neutron guide (blue curve) for wavelength 4 Å. The parameters of the polarizer are shown in Table 2.

The same graphs as in Fig. 15, but in a smaller angular range are shown in Fig. 16 for clarity. The angular distribution of (-) spin component of the beam (green curve), taking into account the reflection from the walls of the neutron guide, is also shown in this figure.

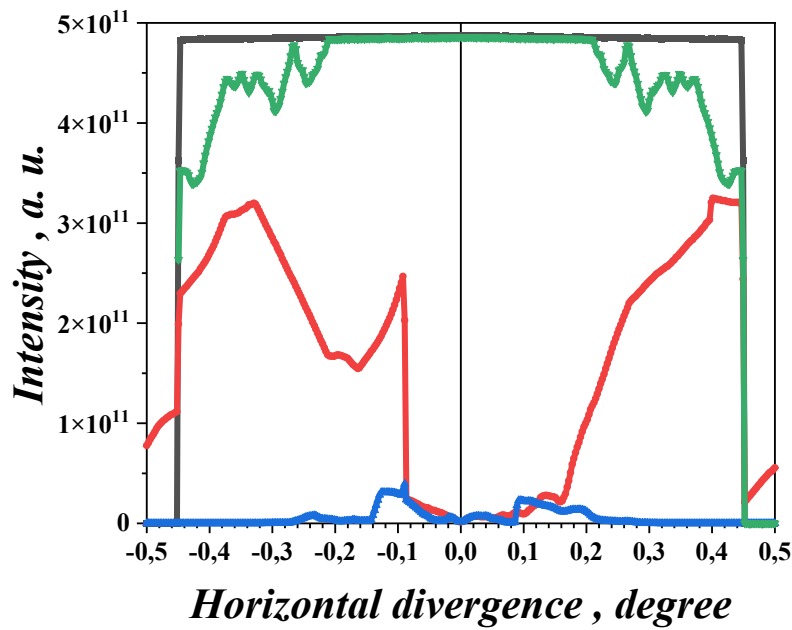

**Fig. 16.** The same graphs as in Fig. 15, but in a smaller angular range for clarity, and the angular distribution of the beam intensity of (-) spin component of the beam (green curve) is also presented.

As can be seen from the graph, for the curve of (+) spin component of the beam at the exit of the polarizer (blue curve), the peaks at large angles disappeared, and near the beam axis they were strongly suppressed. The curve of (-) spin component of the beam (green curve) passes through the entire angular range of the input beam, weakening only at the edges due to the fact that the neutron reflection coefficient from the walls of the neutron guide is less than one. Thus, the angular range of the beam at the exit of the polarizer is increased by 5 times compared to the option of using a single kink. The polarization of the exit beam is high and equal to -0.973. Neutron absorption in all silicon plates and reflection of (-) spin component of the beam from the walls of the kink were not taken into account.

The calculated spectral dependences of the transmission of (-) spin component of the beam $T^-$ and the polarizing efficiency $P$ of the polarizer with the parameters from Table 2, taking into account the absorption of the beam in silicon, are shown in Fig. 17. The angular divergence of the beam at the entrance to the polarizer is ± 0.45 degrees. As follows from the graph, the values of $T^-$ and $P$ are high in the entire spectral range. The transmission of (-) spin component of the beam $T^-$ will increase even more when using thinner silicon wafers with a thickness of 0.15 mm compared to wafers with a thickness of 0.3 mm.

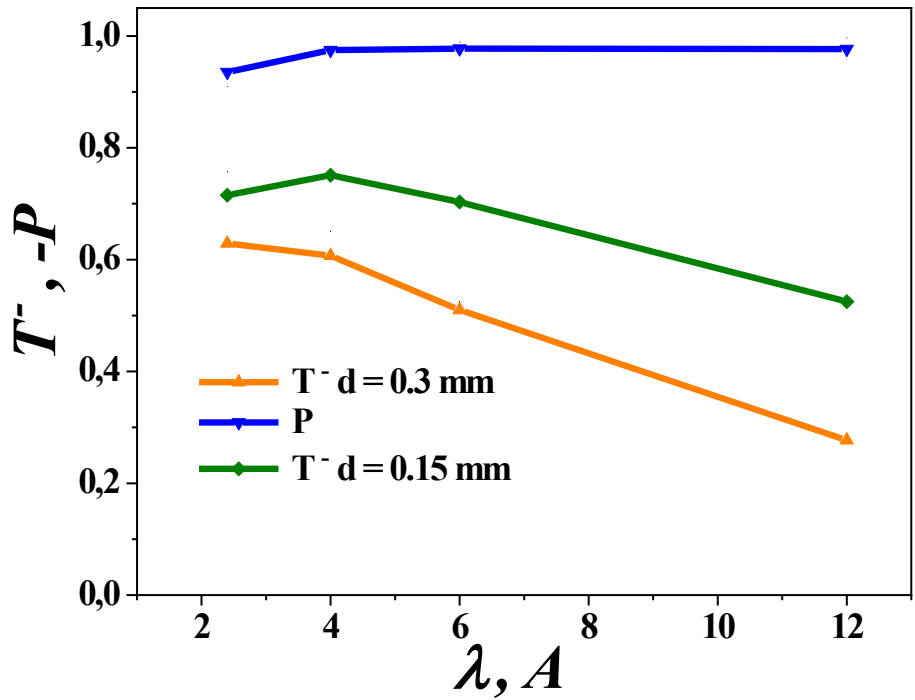

**Fig. 17.** The calculated spectral dependences of the transmission of the (-) spin component of the beam $T^-$ and the polarizing efficiency $P$ of polarizer with the parameters from Table 2, taking into account the absorption of the beam in silicon.

The scheme of spectrometer of inelastic polarized neutron scattering *IN3* is shown in Fig. 18. The main parameters of the *IN3* beam: wavelength range 1.0 – 2.4 Å, beam cross section significant - 60x90 $mm^2$. In *IN3*, it is planned to use the proposed transmission polarizer (item 8).

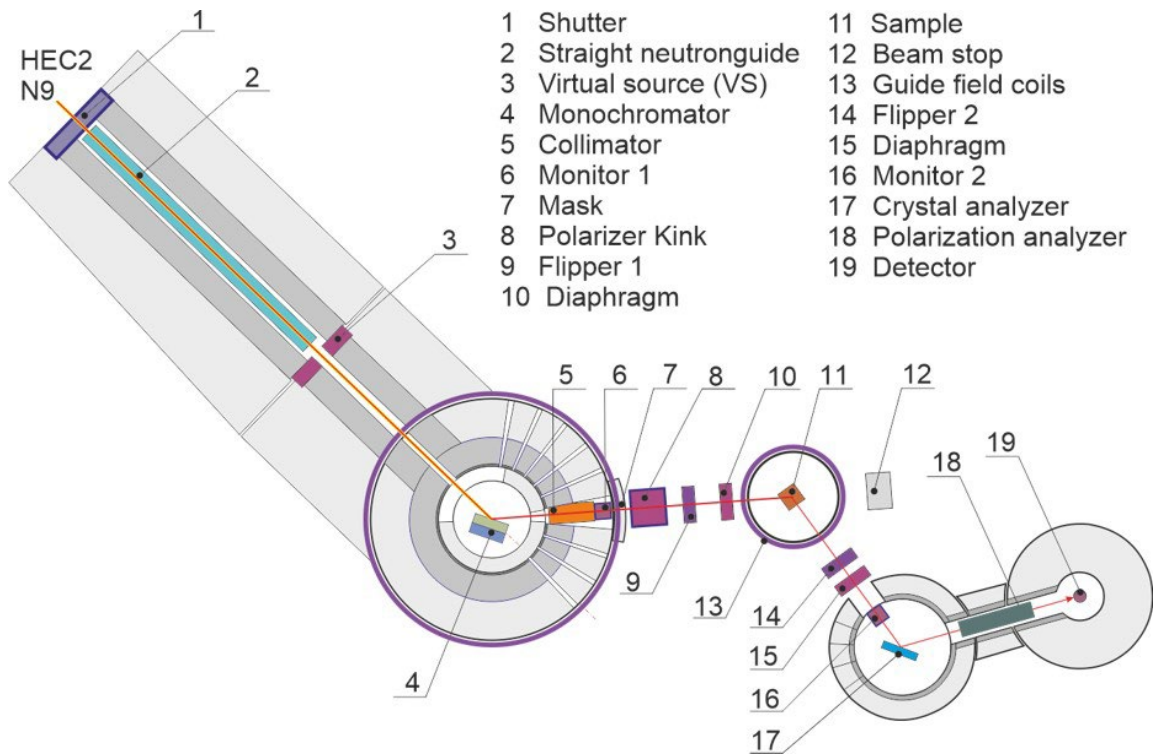

**Fig. 18.** Scheme of spectrometer of inelastic polarized neutron scattering *IN3*.

The parameters of polarizer for *IN3* are shown in Table 3.

**Table 3.** The parameters of polarizer for *IN3*.

| $d$, mm | $L_1$, mm | $L_2$, mm | $L_3$, mm | $\theta_1$, mrad | $\theta_2$, mrad | $m$ |
|---|---|---|---|---|---|---|
| 0.15 | 55+75 | 50 | 144 | 2.8 | 2.0 | 2 |

The calculated graphs of angular intensity distributions for the beam at the entrance (black curve) of polarizer with a divergence of ± 0.20 degrees, for (+) spin component of the beam at the exit from the kink (red curve), for (+) spin component of the beam at the exit from the straight polarizing neutron guide (blue curve) and for (-) spin component of the beam at the exit of the neutron guide (green curve) for a wavelength of 1.5 Å for the polarizer with the parameters from Table 3 are shown in Fig. 19. These curves are similar to the curves shown in Fig. 15 and 16 for the polarizer *IN2* for a wavelength of 4 Å.

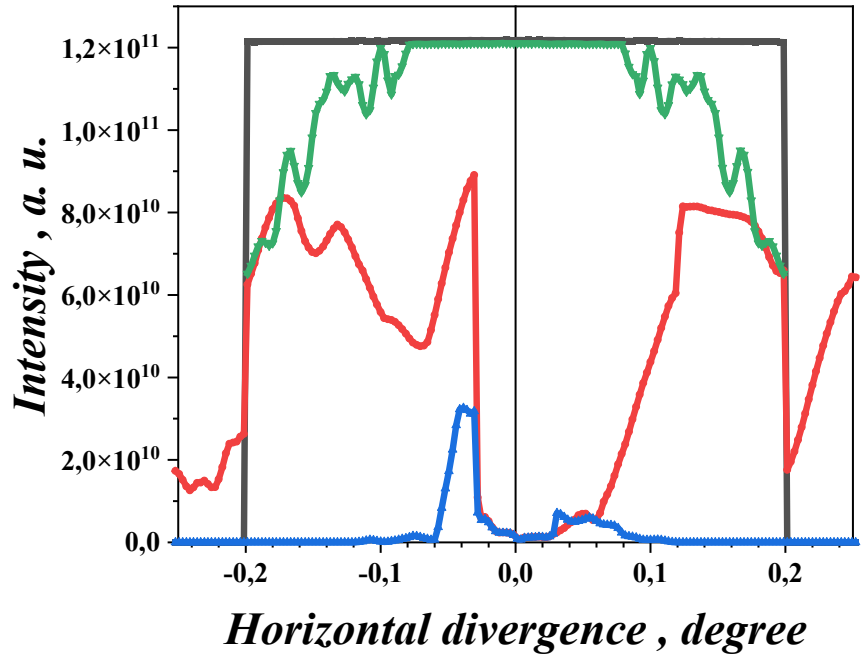

**Fig. 19.** The calculated graphs of angular intensity distributions for the beam at the entrance (black curve) of polarizer with a divergence of ± 0.20 degrees, for (+) spin component of the beam at the exit from the kink (red curve), for (+) spin component of the beam at the exit from the straight polarizing neutron guide (blue curve) and for (-) spin component of the beam at the output of the straight neutron guide (green curve) for a wavelength of 1.5 Å for a polarizer with the parameters from Table 3.

The calculated spectral dependences of the transmission (-) spin component of the beam $T^-$ and the polarizing efficiency $P$ of polarizer, taking into account the absorption of the beam in silicon for the *IN3* polarizer with the parameters from Table 3 are shown in Fig. 20. The angular divergence of the beam at the entrance to the polarizer is ± 0.20 degrees. As follows from the graph, the values of $T^-$ and $P$ are high in the entire spectral range.

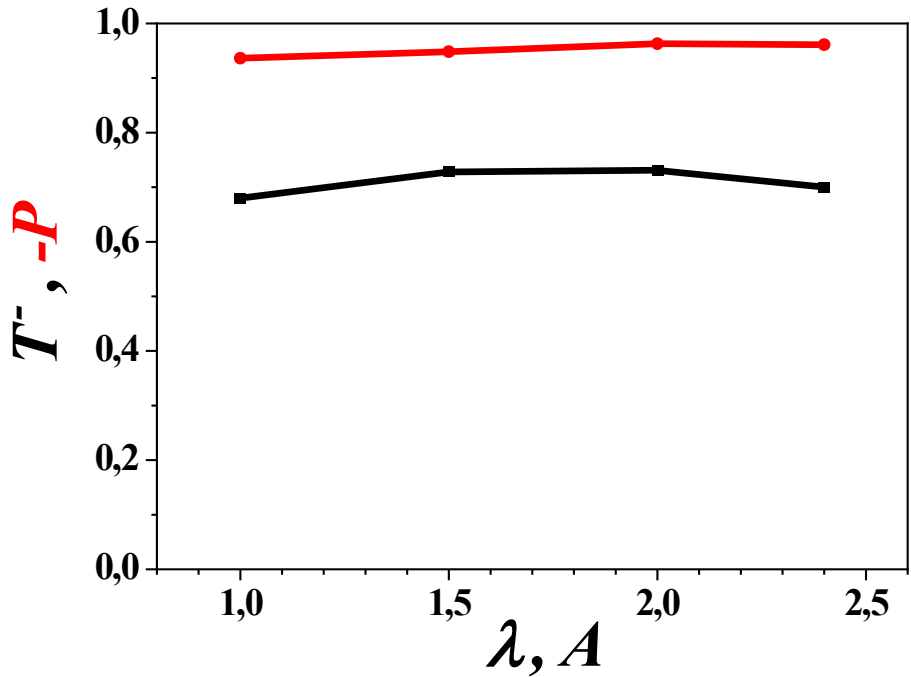

**Fig. 20.** The calculated spectral dependences of the beam transmission (-) spin component of the beam $T^-$ and the polarizing efficiency $P$, taking into account the absorption of the beam in silicon for the *IN3* polarizer with the parameters from Table 3. The angular divergence of the beam at the entrance to the polarizer ± 0.20 degrees.

It should be noted that the use of polarizing supermirror coatings with a higher value of the parameter *m* in the proposed polarizer will significantly reduce the total length of the polarizer and increase the transmission coefficient of the (-) spin component of the neutron beam through the polarizer due to a decrease in absorption in silicon. For example, using supermirrors with parameter $m = 5$ instead of supermirrors with parameter $m = 2$ with a thickness of 0.3 mm silicon wafers will reduce the total length of the kink and the straight polarizing neutron guide of the polarizer by 2.5 times from 225 mm (Table 2) to 90 mm! In addition, the angular width of the beam passed through the polarizer will noticeably increase.

The method of alternating silicon wafers coated on both sides with gadolinium oxide with air gaps of the same size as the wafers is used in solid-state Soller collimators from SwissNeutronics [3] to improve the transmission of a neutron beam through these collimators. It is possible to consider as an option the same method of alternating plates for the elements of this polarizer.

The kink scheme, in which plates polished on both sides from a weakly absorbing material (M) (for example, silicon) alternate with air gaps (Air) of the same size (d) as the plates is shown in Fig. 21. A supermirror coating (SM) is evaporated to each side of the plate. In this case, the attenuation of the beam passing through the kink due to its absorption in silicon will be reduced. In addition, an improvement in the assembly performance of the kink plates is possible in this case. But, for neutrons of the (-) spin component of the beam passing through the air channel, it is now impossible to neglect the reflection at the "air – supermirror" boundary. In this case, these neutrons, whose glancing angle will not exceed the value of the critical angle for a given boundary, will deviate from their original trajectories. Neutrons of the (-) spin component of the beam passing through the silicon channel at the "supermirror - air" boundary will also deviate from their initial trajectories, but already in connection with refraction. All this can lead to a slight decrease in the neutron transmission coefficient of this spin component through a direct polarizing neutron guide. Therefore, an additional calculation of the transmission of the (-) spin component of the beam through a polarizer with such a kink scheme will be required.

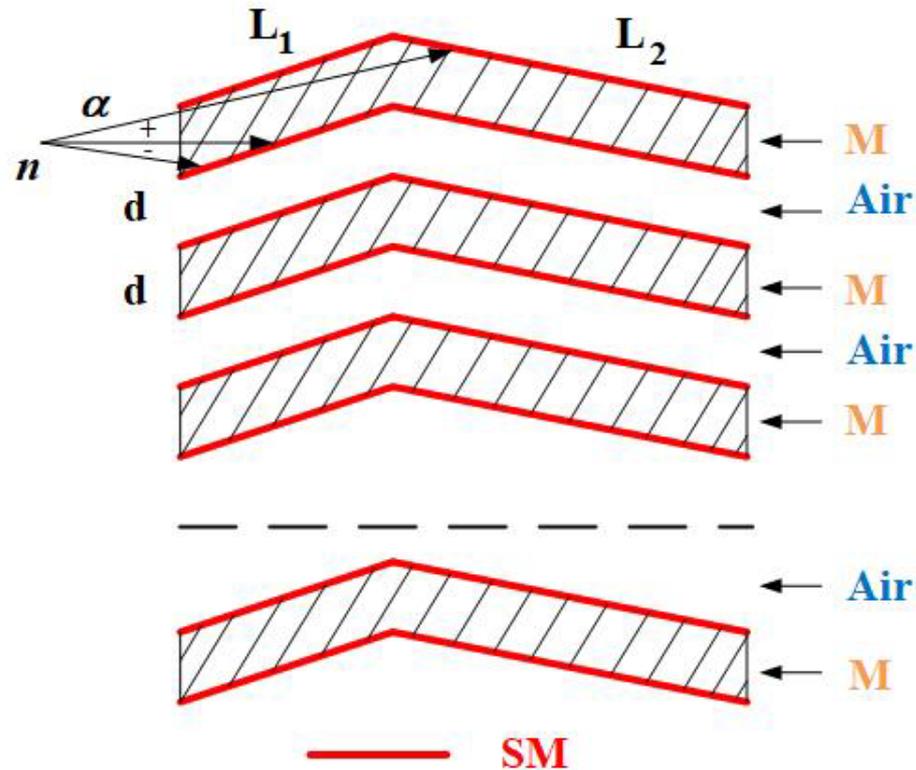

**Fig. 21.** The kink scheme, in which plates made of a weakly absorbing material (M) (for example, silicon) alternate with air gaps (Air) of the same size (d) as the plates.

A variant of the scheme of a straight polarizing neutron guide, which is part of the polarizer, in which plates made of a weakly absorbing material (M) (for example, silicon) alternate with air gaps of the same size (d) as the plates is shown in Fig. 22. Supermirror coating (SM), absorbing layer (AL) and again, the supermirror coating (SM) is evaporated sequentially to each side of the plate. In this case, the attenuation of the beam passing through the neutron guide due to its absorption in silicon will be reduced. In addition, an improvement in the assembly performance of the straight neutron guide plates is possible in this case.

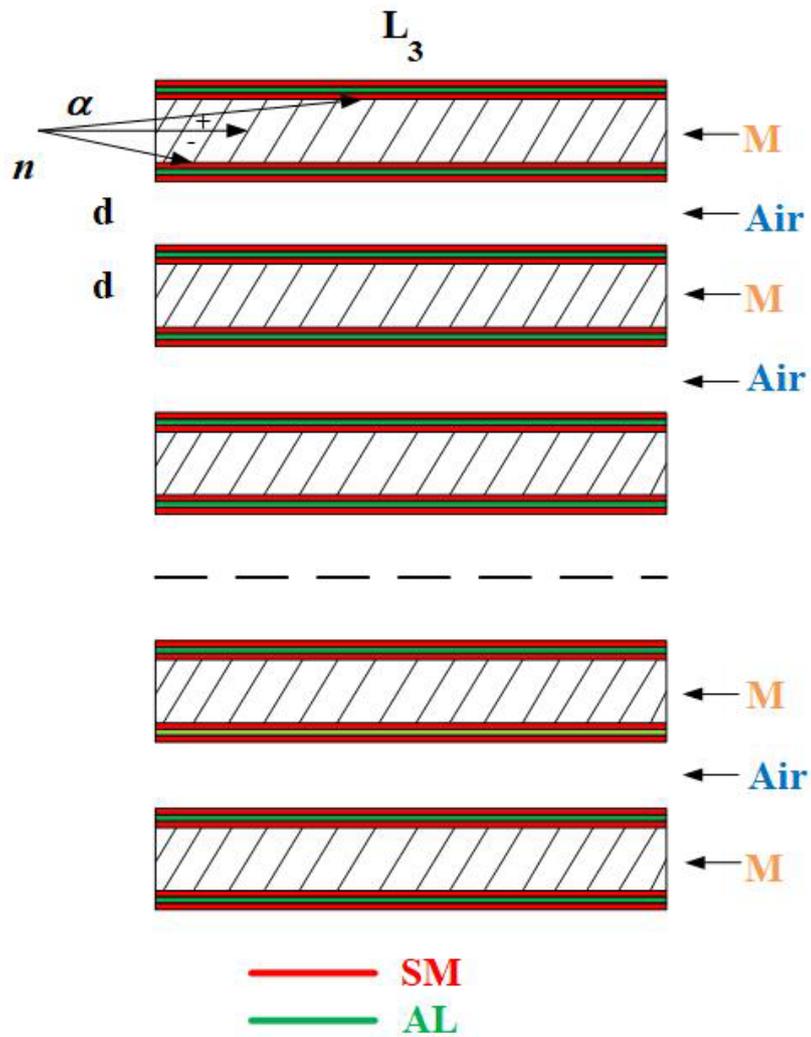

**Fig. 22.** The straight polarizing neutron guide scheme, which is part of the polarizer, in which plates made of a weakly absorbing material (M) (for example, silicon) alternate with air gaps (Air) of the same size (d) as the plates.

# Conclusions

I.      A new neutron supermirror multichannel solid-state transmission polarizer - **TRUNPOSS** (**T**ransmission **Ru**ssian **N**eutron **Po**larized **S**upermirror **S**ystem) is considered.

II.     The main features of the new polarizer:

1. Compact.

2. The polarizer does not increase the angular and spatial distribution of the output beam.

3. Wide angular distribution is available for this polarizer.

4. The beam at the exit has the same direction as at the entrance.

5. The output beam has high polarizing efficiency and luminous intensity.

III.    The high efficiency of using this polarizer for a number of neutron physics facilities of the PIK reactor is shown: *IN2, IN3, DEDM, SEM, TENZOR*.

IV.     The use of polarizing supermirror coatings with a parameter $m > 2$ will significantly increase the efficiency of the proposed polarizer.

V. A detailed consideration of the use of an air gap polarizer between plates in the kink and straight polarizing neutron guide is also of interest to increase the efficiency of the proposed polarizer.

# Acknowledgements

The author considers it his pleasant duty to thank A.G. Pshenichnaya for calculations of the transmission of a neutron flux through the proposed polarizer, Dr. I.A. Zobkalo, Dr. A.N. Matveeva, Dr. S.Yu. Semenikhin, A.P. Bulkin, K.Yu. Terentyev, O.V. Usmanov, M.V. Lasitsa for their interest in the work, and also Dr. Michael Schneider (SwissNeutronics) for his interest in the work and the proposal to use alternating silicon wafers with air gaps in the kink design, as is done in the Soller collimators of SwissNeutronics.